\documentclass[12pt]{iopart}
      \newcommand{\beq}{\begin{equation}}
      \newcommand{\eeq}{\end{equation}}
      \newcommand{\beqa}{\begin{eqnarray}}
      \newcommand{\eeqa}{\end{eqnarray}}
      
      \newcommand{\nn}{\nonumber}

      \newcommand{\del}{\partial}

    \renewcommand{\(}{\left(}
    \renewcommand{\)}{\right)}
      \newcommand{\al}{\alpha}

      \newcommand{\de}{\delta}
      \newcommand{\ep}{\epsilon}

      \newcommand{\la}{\lambda}
      \newcommand{\si}{\sigma}
      
      \newcommand{\La}{\Lambda}
      
      \newcommand{\ba}{\mbox{\boldmath $a$}}    
      \newcommand{\bb}{\mbox{\boldmath $b$}}
     \newcommand{\bg}{\mbox{\boldmath $g$}}
     \newcommand{\bc}{\mbox{\boldmath $c$}}
     
     \newcommand{\bee}{\mbox{\boldmath $e$}}
     \newcommand{\bef}{\mbox{\boldmath $f$}}
     \newcommand{\bh}{\mbox{\boldmath $h$}}
     \newcommand{\bbs}{\mbox{\boldmath $s$}}
     \newcommand{\bu}{\mbox{\boldmath $u$}}
     \newcommand{\bv}{\mbox{\boldmath $v$}}

        \newcommand{\bp}{\mbox{\boldmath $p$}}
        
    \newcommand{\bx}{\mbox{\boldmath $x$}}

    \newcommand{\bV}{\mbox{\boldmath $V$}}
    \newcommand{\bF}{\mbox{\boldmath $F$}}
    \newcommand{\bG}{\mbox{\boldmath $G$}}
    \newcommand{\bH}{\mbox{\boldmath $H$}}
    \newcommand{\bR}{\mbox{\boldmath $R$}}
     \newcommand{\bX}{\mbox{\boldmath $X$}}    
     \newcommand{\bY}{\mbox{\boldmath $Y$}}         
      
       \newcommand{\bzero}{{\bf 0}}

\newcommand{\abs}[1]{\left| #1\right|}

\usepackage{graphicx}
\usepackage{dcolumn}
\usepackage{bm}
\begin{document}
\title[An asymptotic form for marginal running coupling constants]
{An asymptotic formula for marginal running coupling constants and 
 universality of  loglog corrections
    }
 
 \author{Hisamitsu Mukaida}
 \address{Department of Physics, Saitama Medical College, 
 981 Kawakado, Moroyama-cho, Iruma-gun, Saitama, 350-0496, Japan
 }
 \ead{mukaida@saitama-med.ac.jp}

\begin{abstract}
 Given a two-loop beta function for multiple marginal coupling constants, 
 we derive an asymptotic formula for the running coupling constants driven to  
  an infrared fixed point. 
 It can play an important role in universal loglog corrections to physical quantities.
 \end{abstract}
 
\submitto{\JPA}
\pacs{05.10.Cc, 64.60.Ak, 64.60.Fr}

\maketitle
\section{Introduction}
Log and loglog corrections to physical quantities in critical  phenomena generally 
appear in  a statistical system at the  critical dimension.  
In the language of renormalization 
group (RG)  \cite{wk},  those corrections arise from  marginally irrelevant 
 coupling constants in the system.   
 As the length scale we are looking at  becomes larger, the coupling constants 
 effectively change, obeying a renormalization-group equation (RGE),  
 and approach
  an infrared fixed point if initial values of the trajectories are on the critical surface.  
 Universality (i.e., property independent of the initial values)
 of the logarithmic corrections is closely related to the
long-distance   behavior of the running coupling constants. 
 
 Generally, the log and loglog corrections are obtained respectively
 from the leading and the next-to-leading order of the beta function specifying 
  the RGE. 
In order to see this,  it is instructive to consider the case of a single  coupling 
constant.  When the origin is an infrared fixed point, 
an  RGE of a marginal coupling constant $g$ up to the next-to-leading order 
 is generally described as 
 \beq
   \frac{\rmd  g}{\rmd  t} = - a g^2 + b g^3,   \ \  a > 0, 
 \eeq
where $t$ is related  to a length scale $L$  
of RG transformation (RGT) by $t = \ln L$.  
Here we consider the weak-coupling region between the two 
fixed points, 
$0 < g < a/b$. 
It  is  readily integrated and the solution $g(t)$ satisfies  
\beq
  a t + C = \frac{1}{g(t)} - \frac{b}{a} \ln \left[ g(t) \(\frac{a}{b} - g(t)\)^{-1}\right]
\eeq
in this region.  The constant $C$ is determined from an initial condition. 
When $t \rightarrow \infty$, $g(t)$ has the following asymptotic form:  
\beq
  g(t) = \frac{1}{a t} + \frac{b \ln t}{a^3  t^2} + {\rm O}(t^{-2}), 
  \label{single}
\eeq
which  implies  that 
the first term contributes to a $\ln L$ correction, 
while the second term to a $\ln \ln L$ correction. 
 A key feature is that the coefficients of $1/t$ and $\ln t /t^2$ are independent 
of $C$, which leads to  universal logarithmic corrections.  

Although we can integrate RGE explicitly in the case of a single coupling constant, 
we cannot generally perform the same procedure in the case of multiple coupling 
constants.  Therefore,  it is worthwhile to determine an asymptotic form analogous to 
(\ref{single}) in the case of  multiple marginal coupling constants, which 
 is  the main subject of this paper. 

One cannot linearize RGE about a fixed point in the case of 
marginal coupling constants, which complicates the problem finding 
 an asymptotic form without explicit integration.  
An algebraic method was found  in Refs  \cite{im,m}, where 
 the beta function is  restricted to the lowest order. 
Since  the  lowest-order beta function for marginal coupling constants 
is homogeneous, the RGE is invariant under  scaling transformation \cite{im}.  
One can define 
another RG transformation to the RGE,  thanks to the scale invariance.
\footnote{
A general idea of  RG,   applied as a tool for asymptotic analysis 
of  non-linear differential equations,  is developed in 
Refs. \cite{cgo,bk}.  
}
The new RGE generally has a linear term, which allows us to obtain  
 the asymptotic form without explicit integration. 

However,  we cannot apply the above method when higher orders of the 
beta function are taken into account  
because 
there are no such scale invariances.  
Hence we need to find an alternative  method  to remedy the problem 
for linear terms to vanish.   

In the next section,  we present a change of variables in  the RGE 
that allows us to apply the linearization.   
In section \ref{section3},   we  switch 
from the resultant RGE 
 to an equivalent integral equation,  
 and outline the existence of a unique solution 
driven to  the fixed point.  We also give an estimation to the solution.  
Details of the proof are found in  \ref{construction}. 
In section \ref{uaf},  using the estimation found
 in the preceding section,   
we show a sufficient 
condition for loglog corrections  be universal.   
A universal asymptotic formula for the solution in the long-distance limit is 
also derived under  the sufficient condition. 
In section \ref{xyex},  applying our result,   we rederive  
the universal asymptotic formula for the running coupling constants
 in the classical XY model,  as an example.  The result is 
consistent with the original article by  Amit {\it et al.}  \cite{agg}.   
The final section is devoted to summary and discussion.

\section{Changing variables of RGE}
\label{section2}
We consider an RGE for  marginal coupling constants denoted by 
$\bg(t) = (g_1(t), ..., g_n(t))$.  We regard the space of the coupling constants as
the $n$ dimensional Euclidean space $\bR^n$.  
Suppose that we have obtained the RGE up to 
the next-to-leading order, which is to say 
 we start with the following RGE
\beq
  \frac{\rmd  \bg(t)}{\rmd t} = \bV(\bg(t)) + \bF(\bg(t)). 
\label{RGE}
\eeq
The leading and the subleading terms of the beta function are  
 described  by $\bV$ and $\bF$ respectively. 
It is assumed that they possess  the following scaling property: 
\beq
  \bV(k \bg) = k^{2} \bV(\bg), \qquad
  \bF(k \bg) = k^{3} \bF(\bg). 
\label{sp}
\eeq
In general, $\bV$ and $\bF$ are obtained as  quadratic and  cubic polynomials 
in the coupling constants respectively.  
The beta function is defined on the whole space $\bR^n$ 
in this case.  
However, in some cases, the beta function is obtained in a rational form  (e.g., see 
 \cite{xr}).  
For this reason, it is suitable to assume that  $\bV$ and $\bF$ 
are defined on some region $E$ in $\bR^n$.\footnote{
More precisely, $E$ is open subset of $\bR^n$ whose closure contains 
the origin.  We also assume that 
$\bV$ and $\bF$  belong to $C^2\(E\)$, 
 i.e., their second derivatives exist and are continuous on $E$. 
}

It is a general feature of an
RGE  of marginal coupling constants that 
there are no linear terms, which causes 
difficulty in  deriving an asymptotic formula. 
We introduce new variables to bypass this 
problem. 
First we replace $t$ by 
\beq
  u \equiv \frac{1}{\ep} \log \(\ep t + 1\), 
\label{u}
\eeq
where $\ep$ is a parameter with 
\beq
  0 < \ep < 1. 
\eeq
As we will see later, $\ep$ is introduced to control an effect of the subleading term $\bF$. 
Next we  change  $\bg$ to $\bc$, where 
\beq
  \bg(t) = \rme^{- \ep u} \bc(u). 
  \label{bc}
\eeq
The left-hand side of (\ref{RGE}) becomes 
\beqa
   \frac{\rmd  \bg(t)}{\rmd t} &=& \rme^{-\ep u} \frac{\rmd }{\rmd u} \(\rme^{-\ep u} \bc(u)\)
   \nn\\
   &=&  \rme^{-2 \ep u} \( - \ep \bc(u) +  \frac{\rmd \bc(u)}{\rmd u}\), 
\eeqa
while the right-hand side is 
\beq
  \bV(\rme^{-\ep u}\bc)+ \bF(\rme^{-\ep u} \bc) = \rme^{-2 \ep u} \bV(\bc) + \rme^{-3 \ep u}\bF(\bc)
\eeq
because of the scaling property (\ref{sp}). In this way, 
the RGE (\ref{RGE})  is written as\footnote{
We comment that 
this transformation is generalized in the case where a beta function starts with 
an $m$th homogeneous function ($m \geq 2)$.  In fact, if 
 we define $u$ and $\bc(u)$ by 
 $(m-1) \ep u =  \ln(\ep t +1)$ and $\bg(t) = \rme^{-\ep u} \bc(u)$ respectively, 
 we get a similar equation having a linear term.
}
\beq
   \frac{\rmd  \bc(u)}{\rmd  u} =  \ep \bc(u) + \bV(\bc(u)) + \rme^{-\ep u} \bF(\bc(u)). 
\label{RGEa}
\eeq

Now we extract the linear part from
 the first two terms.  We assume that a non-trivial solution 
$\bc^* \in E$ for 
\beq
   \ep \bc^{*} +  \bV\(\bc^{*}\) = {\bf 0}
\label{fxd}  
\eeq
exists.  Note that $\bc^*$ is linear in $\ep$. 
In fact,  let us introduce $\ba^*$ by 
\beq
   \bc^{*}  \equiv \ep \ba^{*}. 
   \label{fxdc0}
\eeq
Because of the scaling 
property (\ref{sp}), $\ba^{*}$ is determined by 
\beq
  \ba^{*} +  \bV\(\ba^{*}\) = {\bf 0}. 
\label{fxdc}
\eeq
Hence $\ba^*$ is independent of $\ep$ and (\ref{fxdc0})  indicates that 
$\bc^*$ approaches the origin as $\ep$ becomes smaller. 
Therefore,  an effect of the subleading term $\bF$ 
in a neighborhood of 
$\bc^*$ is suppressed if  we take $\ep$ sufficiently small. 
It plays an important role in  showing the existence of  
 a solution for (\ref{RGEb})
by a contraction map, as we will see in  \ref{construction}.

We analyze (\ref{RGEa}) in a neighborhood of $\bc^*$.  Define
\beq
  \bb(u) \equiv \bc(u) -\bc^{*} 
\eeq
and write 
\beq
  \bV(\bc(u)) = \bV(\bc^{*} ) + D\bV(\bc^{*}) \bb(u) + \bv(\bb(u)), 
\eeq
where $D\bV(\bc^{*})$ is the derivative of $\bV$ at $\bc^{*}$, which 
is represented by the $n\times n$ matrix as 
\beq
  D\bV(\bc^*)_{ij} = \frac{\del V_i}{\del c_j}(\bc^*). 
\eeq  
The RGE (\ref{RGEa}) is written as 
\beq
   \frac{\rmd  \bb(u)}{\rmd  u} =  M \bb(u) + \bH(u, \bb(u)), 
\label{RGEb}
\eeq
where 
\beq
  M \equiv \ep I_{n} + D\bV\(\bc^{*}\), \qquad
  \bH(u, \bb(u))  \equiv \bv(\bb(u)) + \rme^{-\ep u} \bF(\bc^{*} + \bb(u))
  \label{M}
\eeq
with $I_{n}$ being the $n \times n$ unit matrix.   
Further,  (\ref{sp}) yields\footnote 
{
To prove (\ref{spdv}), we use  (\ref{sp}) again. 
$$
  \bV(\bc^{*} + \bh) = \bV(\bc^{*}) + D\bV(\bc^{*}) \bh + o(\abs{\bh}), 
$$
while the left-hand side is 
$$
  \bV(\bc^{*} + \bh) = \ep^{2}\bV(\ba^{*} + \bh/\ep) = \ep^{2}\bV(\ba^{*}) 
  + \ep D\bV\(\ba^{*}\) \bh + o(\abs{\bh}). 
$$
Comparing the linear term in $\bh$, we have (\ref{spdv}).  
}
\beq
 D\bV\(\bc^{*} \) = \ep D\bV\(\ba^{*}\). 
\label{spdv}
\eeq
Therefore $M$ is linear in $\ep$
\beq
  M = \ep \( I_n + D\bV\(\ba^*\) \). 
  \label{def_m}
\eeq

To sum up, if we know a solution $\bb(u)$ for  (\ref{RGEb}), 
we get a solution for (\ref{RGE}) by 
\beq
  \bg(t) = \rme^{-\ep u} \(\bc^* + \bb\(u\)\), \ \ u = \frac{1}{\ep}\ln\(\ep t + 1\). 
\label{g(t)}
\eeq

\section{Integral equation}
\label{section3}
In this section, following reference \cite{p} (Section 2.7),  we derive 
an integral equation satisfied by  a solution $\bb(u)$ for 
(\ref{RGEb}) driven to $\bzero$. 

We assume that there are no eigenvalues with zero real part in $M$. 
Suppose that $M$ has $k$ eigenvalues having a negative real part,  and 
$n-k$ eigenvalues with positive real part. 
If the positive  eigenmodes are fine-tuned to vanish, $|\bb(u)|$ 
becomes smaller as $u \rightarrow \infty$.  
We find from (\ref{g(t)}) that the corresponding 
$\bg(t)$ approaches the origin from the $\bc^*$-direction. 
In order to show the existence of such solutions,   we 
decompose $M$ into 
a block diagonal form. Namely, 
\beq
  R^{-1}MR = 
  \left(
  \begin{array}{cc}
  \ep P&0\\
  0&\ep Q
  \end{array}
  \right) \equiv \ep \Lambda, 
\eeq
where $\ep P$ is a $k\times k$ matrix whose eigenvalues have  a negative 
real part.  Similarly, $\ep Q$ is an $(n-k) \times (n-k)$ matrix, where 
its eigenvalues have a positive real part.   
Note that $P$, $Q$ and $\La$ are independent of $\ep$ because $M$ is linear in $\ep$.
Define the tilde operation 
\beq
  \tilde{\bx} = R^{-1}\bx, \qquad \tilde{\bX} = R^{-1} \circ \bX \circ R
\eeq
for a point $\bx \in E$ and a map 
$\bX : E \rightarrow \bR^{n}$, e.g., $\tilde{\bF}(\tilde{\bc}) = R^{-1}\bF(R\tilde{\bc}) = R^{-1}\bF(\bc)$. 
The RGE (\ref{RGEb})  can be written as 
\beq
  \frac{\rmd  \tilde{\bb}(u)}{\rmd  u} = \ep \Lambda \tilde{\bb}(u) + \tilde{\bH}(u, \tilde{\bb}(u)). 
\eeq
Let 
\beq
  U(u) \equiv 
    \left(
  \begin{array}{cc}
   \rme^{P \ep u} &0\\
  0&0
  \end{array}
  \right),  
  \qquad
   T(u) \equiv    
   \left(
  \begin{array}{cc}
   0 &0\\
  0& \rme^{Q\ep u}
  \end{array}
  \right). 
\eeq
Then 
\beq
  \frac{\rmd  U}{\rmd  u} = \ep  \Lambda U(u), \qquad
  \frac{\rmd  T}{\rmd  u} =  \ep \Lambda T(u)
\eeq
and 
\beq
  \rme^{\ep\Lambda u} = U(u) + T(u).
\eeq
We focus on  a solution that behaves as $\tilde{\bb}(u) \rightarrow \bzero$ as 
$u \rightarrow \infty$.  
The integral equation corresponding to it is 
 \beq
 \fl
   \tilde{\bb}(u) = U(u) \bp + \int_{0}^{u}\rmd u' U(u-u') \tilde{\bH}(u', \tilde{\bb}(u')) 
   - \int_{u}^{\infty}\rmd u' T(u-u') \tilde{\bH}(u', \tilde{\bb}(u')), 
\label{ie}
 \eeq
where $\bp=(p_1, ..., p_k, 0, ..., 0)$ specifies an initial condition 
in the following way: 
\beqa
  &&\tilde{b}(0)_{i} = p_i \quad {\rm for } \quad i = 1, ..., k. 
  \nn\\
  &&\tilde{b}(0)_{i} = - \( \int_{0}^{\infty}\rmd u' T(-u') \tilde{\bH}(u', \tilde{\bb}(u'))\)_{i}
  \quad {\rm for } \quad i = k+1, ..., n. 
\eeqa 

We can 
show that ({\ref{ie}) has a unique solution if $\ep$ and $\bp$ are sufficiently small. 
Moreover, 
we find that the solution satisfies
\beq
   \left| \tilde{\bb}(u) \right| \leq  J  \rme^{-\al \ep u}
   \label{estb}
\eeq
for some $J>0$. Here $\al$ is a positive number,  
such that $-\al$ is strictly greater than the real part of 
every eigenvalue of $P$.   In order to prove (\ref{estb}), 
we need to extend the usual proof of 
the stable manifold theorem (see  \cite{p}), 
which is applied to the case of an autonomous system, 
to the case of the nonautonomous system (\ref{RGEb}).  
 Details of a proof of (\ref{estb}) are 
found in    \ref{construction}.

\section{Universal Asymptotic Form of  $\tilde{\bb}(u)$}
\label{uaf}
In this section, we derive a universal asymptotic form of  
 $\tilde{\bb}(u)$ by  applying  (\ref{estb}) to 
the right-hand side in  (\ref{ie}). 
(Here, ``universal'' means that the asymptotic form is independent of 
$\bp$. )

For this purpose, we give  
a more concrete form of $P$.
Let $\la_l$ ($l=1,...,n_-$) be the distinct eigenvalues of $P$ with 
the multiplicity $d_l$. 
We denote by  $W_l$ the generalized eigenspace associated with $\la_l$. 
Clearly, $\dim W_l = d_l$.   Taking an appropriate basis for $\bR^n$, 
$P$ is represented as a block diagonal form.  Here 
the $l$ th block $P_l$ is a $d_l \times d_l$ upper triangle matrix whose 
diagonal components take a common value $\la_l$.   
Furthermore,  the basis allows us to assume that 
 $N_l \equiv P_l - \la_l I_{d_l}$ is a nilpotent matrix, namely 
 $N^{\nu_l-1} \neq 0$, $N^{\nu_l} =  0$ for some $1 \leq \nu_l \leq d_l$. 
An arbitrary element $\bx \in \bR^n$ can be decomposed as 
 \beq
   \bx = \sum_{l=1}^{n_-} \bx^{(l)} +\bx^{(+)},    \ \ 
    \bx^{(l)} \in W_l, 
 \eeq
 where $\bx^{(+)}$ is an element of the subspace spanned by the 
 positive eigenmodes of $M$. 
Applying $U(u)$ to the both sides, we have
\beq
  U(u) \bx = \sum_{l=1}^{n_-}\rme^{\ep \la_l u} 
  \sum_{k=0}^{\nu_l -1} \frac{\(\ep u\)^k N_l^k}{k!} \bx^{(l)}, 
\label{decom} 
\eeq
with the convention $N_l^0 = I_{d_l}$ even if $N_l$ is the zero matrix. 

We can show that\footnote{
In order to prove (\ref{-ep}),  note that
$$
   \bV\(\bc^{*} + h \bc^{*}\) = \bV\(\bc^{*}\) + D\bV\(\bc^{*}\) h \bc^{*} + o(\abs{h}). 
$$
On the other hand, the left-hand side is equal to 
$$
  (1+h)^{2} \bV\(\bc^{*}\) = - \ep   (1+h)^{2} \bc^{*}
$$
because of (\ref{sp}) and (\ref{fxd}).  Comparing the linear term in $h$, 
we get (\ref{-ep}). 
}
\beq
  M \bc^{*} = - \ep \bc^{*}, 
\label{-ep}
\eeq
which implies
$
  P  \bc^{*} = -  \bc^{*}. 
$
Then we set 
\beq
\la_1 = -1. 
\label{-1}
\eeq

Using  (\ref{decom}) and (\ref{-1}),  we can estimate
  the right-hand side of (\ref{ie}).  The first term is written as 
\beq
  U(u) \bp = 
  \sum_{l=1}^{n_-} \bG_1^{(l)}(u)\rme^{\la_l \ep u}, 
\label{ub}
\eeq
where $\bG_1^{(l)}(u)$ is a  polynomial of  degree  at most $\nu_l -1$. 
Since (\ref{ub}) explicitly depends on $\bp$, 
it is non-universal.  

In order to obtain a universal asymptotic form, 
 universal terms  should dominate over (\ref{ub}) 
 when $u\rightarrow \infty$.
Let us find  
a condition that such terms appear from  the 
remaining part. 
The integral containing $U$ in (\ref{ie}) is divided  as  
\beq
\fl
  \int_0^u U(u-u')  
  \rme^{-\ep u'}
  \tilde{\bF}\(\tilde{\bc}^*\) \rmd u'
+
  \int_0^u U(u-u')  
 \( \tilde{\bH}(u', \tilde{\bb}(u')) - \rme^{-\ep u'}
  \tilde{\bF}\(\tilde{\bc}^*\) \)\rmd u'. 
\label{part}
\eeq
Applying  (\ref{decom}) to $\tilde{\bF}(\tilde{\bc}^*)$,  the first integral 
is easily calculated.  It is important to notice that 
the case of $l=1$ has to be  treated separately,  because 
the  factor $\exp(-\ep \la_1 u')$  in $U(u-u')$ cancels 
$\exp(-\ep u')$ in the integrand.   When $l=1$ and $k=\nu_1-1$ in (\ref{decom}), 
the cancellation 
 brings a term proportional to $u^{\nu_1}\exp{(-\ep u)}$, which
  is not contained in $\bG_1^{(l)}(u)$. 
 Writing this  explicitly,  
the integral is expressed as 
\beq
\fl
  \int_0^u U(u-u')  \rme^{-\ep u'} 
  \tilde{\bF}\(\tilde{\bc}^*\) \rmd u'  =   
  \frac{\(\ep u\)^{\nu_1} \rme^{-\ep u}}{\ep \nu_1!}N_1^{\nu_1-1} 
  \tilde{\bF}^{(1)}\(\tilde{\bc}^*\) + \sum_{l=1}^{n_-} \rme^{\la_l \ep u} \bG^{(l)}_2(u). 
\label{intl}
\eeq
Here, it is straightforward to check that 
 $\bG^{(l)}_2(u)$ is a polynomial whose degree is at most $\nu_l-1$, the same order as 
 (\ref{ub}), so that its universal behavior is obscured by the non-universal nature 
of $U(u) \bp$.  
On the other hand, the first term
can dominate over (\ref{ub})  in the case when 
\beq
  \Re\la_l < -1 \ \ \(l = 2, ..., n_-\). 
\label{ess}
\eeq
Moreover,  as we see in  \ref{appc}, this term is most dominant 
 in the right-hand side of 
(\ref{ie}) when (\ref{ess}) holds. Thus,  we obtain 
\beq
   \tilde{\bb}(u) = \frac{\(\ep u\)^{\nu_1} \rme^{-\ep u}}{\ep \nu_1!}N_1^{\nu_1-1} 
  \tilde{\bF}^{(1)} \(\tilde{\bc}^*\) + {\rm O}\(u^{\nu_1 -1} \rme^{-\ep u}\). 
\label{premain}
\eeq
 Let us revert to the original variables by (\ref{g(t)}).  
 We get, under the condition (\ref{ess}), 
\beqa
\fl
  \tilde{\bg}(t) 
  = \frac{1}{\ep t + 1} \( \ep \tilde{\ba}^* +  \frac{\ep^3\(\ln\(\ep t +1\) \)^{ \nu_1}}
  {\ep \(\ep t + 1\)\nu_1!}N_1^{\nu_1-1} 
  \tilde{\bF}^{(1)} \(\tilde{\ba}^*\)\) + {\rm O}\(\frac{\(\ln \(\ep t+1\)\)^{\nu_1-1}}{\(\ep t +1\)^2}\) 
  \nn\\
  \lo=  \frac{\tilde{\ba}^*}{t} +  \frac{\(\ln t \)^{ \nu_1}}
  { t^2 \nu_1!}N_1^{\nu_1-1} 
  \tilde{\bF}^{(1)} \(\tilde{\ba}^*\) + {\rm O}\(\frac{\(\ln t\)^{\nu_1-1}}{t^2}\), 
  \label{res}
  \eeqa
  as $t \rightarrow \infty$.   This is the main result of this paper. 
  It is worthwhile to note that the $(\ln t)^{\nu_1}$ term 
  appears, which brings about a $(\ln \ln L)^{\nu_1}$ correction in general.

  \paragraph{The case of $dim \, W=1$}
 The simplest case is that $\dim W_1 = 1$. 
   In this case, using the unit eigenvector 
   $\tilde{\bee}^* \equiv \tilde{\ba}^*/\abs{\tilde{\ba}^*}$ in $W_1$,  
 \beqa
   \tilde{\bF}^{(1)}(\tilde{\ba}^*)  &=&  \(\tilde{\bF}(\tilde{\ba}^*),  \tilde{\bee}^*\) \tilde{\bee}^*
   \nn\\
   N_1 &=& 0
   \nn\\
   \nu_1 &=& 1. 
 \eeqa
Then the result (\ref{res}) is simplified to 
 \beq
   \bg(t) =  \frac{\ba^*}{t} +  \frac{\ln t}{t^2}
 \(\tilde{\bF}(\tilde{\ba}^*),  \tilde{\bee}^*\) R \tilde{\bee}^*+ {\rm O}\(\frac{1}{t^2}\). 
 \label{cor}
 \eeq
 Note that we have removed '$\tilde{\ }$' from $\tilde{\bg}$ and 
 $\tilde{\ba}^*$ by applying $R$.   Now 
 let us write $R$ as  a set of column vectors
 \beq
   R = \( \bv_1, ..., \bv_n \). 
 \eeq
 Similarly, we write $R^{-1}$ in terms of  a set of row vectors
 \beq
  R^{-1} =  \(
     \begin{array}{c}
     \bu_1\\
     \bu_2\\
     \vdots\\
     \bu_n
     \end{array}
     \). 
 \eeq
 Since $R^{-1}R=I_n$, we have
 \beq
   \(\bv_i, \bu_j\) = \de_{ij}. 
 \eeq
 We know that $\ba^*$ is an eigenvector with the eigenvalue $-1$, so  
 we can take $\bv_1 = \ba^*$.  In this case,  
 \beq
   \tilde{\ba}^* = R^{-1} \ba^* = \(
     \begin{array}{c}
     1\\
     0\\
    \vdots \\
     0
     \end{array}
     \) = \tilde{\bee}^*. 
 \eeq
 Therefore
 \beq
   \(\tilde{\bF}(\tilde{\ba}^*),  \tilde{\bee}^*\) = \tilde{F}_1(\tilde{\ba}^*), 
 \eeq
where the right-hand side is the first component of $\tilde{\bF}(\tilde{\ba}^*)$. 
 Using this convention, the asymptotic behavior (\ref{cor}) is  
 simplified further to 
 \beq
   \bg(t) =  \frac{\ba^*}{t} +  \frac{\ln t}{t^2}
\tilde{F}_1(\tilde{\ba}^*) \ba^*+ {\rm O}\(\frac{1}{t^2}\). 
 \label{cor2}
 \eeq
\section{Application to the two-dimensional XY model}
\label{xyex}
In this section, we illustrate our method using the two-dimensional
 classical XY model \cite{bkt}. 
  The  beta function up to subleading order of this model 
 and the two-point correlation function containing loglog correction 
are originally derived by  Amit \etal.  \cite{agg}.  They  obtained 
the asymptotic form of the coupling constants by explicitly integrating 
the RGE.  Here we rederive the asymptotic form within our formulation.

The 2D classical XY model has the following RGE  \cite{agg,bh}:
\beqa
  &&\frac{\rmd  g_{1}}{\rmd  t} = - g_{2}^{2}-B_{1}g_{2}^{2}g_{1}
    \nn\\
  &&\frac{\rmd  g_{2}}{\rmd  t} = -g_{1} g_{2} - A_{1} g_{2}^{3}, 
  \label{xy}
\eeqa
where $g_2 > 0$ and $2 A_1 + B_1 = 3/2$.  
Thus  
\beqa
   \bV(\bg) = 
  \left(
  \begin{array}{c} 
  -g_2^2 \\
  -g_1 g_2
  \end{array}
  \right), 
\ \ 
  \bF(\bg) = 
  \left(
  \begin{array}{c} 
  -B_{1} g_{2}^{2} g_{1} \\
  -A_{1}g_{2}^{3}
  \end{array}
  \right). 
  \eeqa
Solving (\ref{fxdc}), 
we get a non-trivial solution 
\beq
  \ba^{*} = {1 \choose 1}. 
\eeq
Inserting this into (\ref{def_m}), one finds 
\beq
  M =
  \left(
           \begin{array}{cc}
            1 & -2 \\
            -1& 0 \\
            \end{array}
    \right) \ep. \ \ 
\eeq
The eigenvalues and corresponding eigenvectors  of $M$ are
\beq
  -1 \leftrightarrow {1 \choose 1}  \ \ {\rm and} \ \ 2 \leftrightarrow {-2 \choose 1}. 
\eeq
Namely, the space of negative eigenmodes of $M$ is one-dimensional. 
It indicates  that the critical surface along 
$\ba^*$ is in fact  a line.   
The transformation matrix $R$ and the diagonalized matrix $\La$ are 
 obtained from the eigenvectors and the eigenvalues respectively.  
 The result is 
\beq
     R = 
     \left(
           \begin{array}{cc}
            1 & -2 \\
            1 & 1 \\
            \end{array}
    \right),  \qquad 
       \Lambda = 
     \left(
           \begin{array}{cc}
            -1 & 0 \\
            0 & 2 \\
            \end{array}
    \right). 
\eeq
The condition (\ref{ess}) is satisfied and 
 $\dim W_1 =1$ in this example.   
 Furthermore, since we chose 
 $\ba^*$ as the first column in $R$, 
 (\ref{cor2}) is applicable.  
Using $R$, we compute 
\beqa
%
\tilde{\bF}(\tilde{\ba}^{*})= R^{-1} \bF(\ba^{*}) = 
  -\frac{1}{3}\left(
  \begin{array}{c} 
  2A_{1}+B_{1}\\
  A_{1}-B_{1}
  \end{array}
  \right). 
  \eeqa
From the first component of the above result,  we conclude that
\beqa
  \bg(t) &=& \frac{\ba^{*}}{t} -\frac{1}{3}\(2A_{1} + B_{1} \) \frac{\log t}{t^{2}} \ba^{*}
  + 
  {\rm O}\( \frac{1}{t^{2}} \)
  \nn\\
  &=&  \frac{\ba^{*}}{t} -\frac{1}{2}  \frac{\log t}{t^{2}} \ba^{*}
  + 
  {\rm O}\( \frac{1}{t^{2}} \)
 \label{asymxy}
\eeqa
for the critical line.   This is consistent with 
the original result. 

\section{Summary and Discussion}
We have obtained an asymptotic formula for multiple marginally 
irrelevant coupling 
constants in the case where the two-loop beta functions are known. 
We first change the variables in the given RGE as (\ref{u}) and (\ref{bc}). 
One can extract a linear part $M$ defined  in (\ref{M}) 
from the resulting differential equation 
if  there is a real solution $\bc^*$ for  (\ref{fxd}), 
 although the original RGE cannot  be linearized.  
 It is assumed that there are no eigenvalues of $M$ 
 with zero real part in the present investigation. 

Next we have shown that, if we take $\ep$ in (\ref{bc}) to be 
sufficiently small, 
there is a $k$ dimensional neighborhood $N$ of $\bc^*$ such 
that trajectories of the RGE starting in $N$ approach the origin 
along $\bc^*$, where $k$ is  the number of eigenvalues of $M$ 
with a negative real part.  
Furthermore, if the eigenvalues of $M$ with 
a negative real part satisfy  the condition (\ref{ess}), 
 the asymptotic formula 
of $\bg(t)$ becomes universal and is given by (\ref{res}). 

As we have concretely shown by  using 
the two-dimensional classical XY model,  
the advantage of this formula is that we do not need to integrate 
the RGE explicitly.   

The non-linearity having the original RGE is 
changed into solving (\ref{fxd}) in our formalism. 
All quantities appearing in  (\ref{res}) can be 
computed using simple linear algebra.  
Thus,  our approach can be applicable 
even though the RGE is too 
complicated to integrate.   Application to 
 such a complicated  RGE is a future problem. 

Finally,  we comment on  our 
previous investigation on this topic.  
As we mentioned 
briefly in the introduction,  another RGE is derived 
with respect to the scaling invariance possessed 
by the leading-order RGE in reference  \cite{im}, 
where  the asymptotic formula in the 
in the lowest order is obtained.  
It is consistent with the present work.   
The relationship between 
 the linear part of the new RGE and $M$ in (\ref{M}) 
 is clarified in reference \cite{m}.  However,   a direct 
 relationship between $M$ and  the original RGE was not clear. 
In this paper,  it is found that $M$ naturally comes out 
in the RGE by the change of variables, which makes it possible 
to derive the asymptotic formula for a  beta function up to the subleading order. 
The previous formulation can also be applied to deriving
the correlation-length exponent of  phase transitions 
in infinite order, which will be extended to the case of  the
higher-order beta function. 
\\

\ack
It is my pleasure to thank C. Itoi for informing me 
 of this attractive problem.

\appendix
\section{Existence of the solution for (\ref{ie})}
\label{construction}
The purpose of this appendix is to show existence of a unique solution for 
the integral equation 
(\ref{ie}) and derive the estimation (\ref{estb}). 
To this end, we need estimations of $\tilde{\bH}$, $U$ and $T$. 
Throughout this appendix, we omit "$\tilde{\ }$" for brevity, (e.g., 
we write $\bb$ or $\bH$ instead of $\tilde{\bb}$ or $\tilde{\bH}$ 
respectively. )

\subsection{Lipschitz-type condition for $\bv$ and $\bef$}
We first derive a Lipschitz-type condition for $\bv$ and $\bef$ defined through 
 the following equation:  
\beqa
  \bV\(\bc^* + \bb\) &=&  \bV\(\bc^*\) + D\bV\(\bc^*\)\bb + \bv(\bb)
    \label{v0}
\\
  \bF(\bc^* + \bb) &=& \bF(\ba^{*}) + D\bF(\ba^{*}) \bb + \bef(\bb), 
  \label{f}
\eeqa
which is employed for the estimation of $\bH$. 
Let 
\beq
  N_{\eta} \equiv \left\{ \bb;  \bb \in \bR^n, \abs{\bb} < \eta  \right\}. 
\eeq

First we prove that:  
for any  $\xi_1, \xi_2>0$, 
there exists a number $\eta>0$ such that
\beqa
  \abs{\bv \(\bb_1\) - \bv\(\bb_2\)} &<& \xi_1 \ep \abs{\bb_1-\bb_2}
  \label{vandf1}
  \\
 \abs{\bef \(\bb_1\) - \bef \(\bb_2\)} 
 &<& \xi_2 \ep \abs{\bb_1-\bb_2}
 \label{vandf2}
\eeqa
for  $\bb_1, \bb_2  \in  N_{\ep \eta} $ and $0<\ep<1$. 
 
\noindent
{\it Proof}: 
We define $\bv_0$ by the following equation: 
\beq
  \bV\(\ba^* + \bb'\) =  \bV\(\ba^*\) + D\bV\(\ba^*\)\bb' + \bv_0(\bb'). 
\eeq
 For $\bb'_1$, $\bb'_2 $,  let
$\bbs \equiv \bb'_2 - \bb'_1$ and define 
\beq
  \bY(\theta) \equiv \bv_0\(\bb'_1 + \theta \bbs\). 
\eeq
Note that $\bY(0) = \bv_0(\bb'_1)$ and $\bY(1) = \bv_0(\bb'_2)$. 
Taking the derivative with respect to $\theta$, we have
\beq
   \frac{\rmd \bY(\theta)}{\rmd  \theta} = D\bv_0\(\bb'_1 + \theta \bbs\)\bbs. 
\eeq
Integrating from 0 to 1, we get 
\beq
  \abs{\bv_0\(\bb'_1\) - \bv_0\(\bb'_2\)} = \abs{\bY(1) - \bY(0)} \leq
   \int_0^1 \rmd\theta  \abs{D\bv_0\(\bb'_1 + \theta \bbs\)}\abs{\bbs}.
 \label{Y} 
\eeq
Since $D\bv_0$ is continuous and  $D\bv_0\(\bzero\) = \bzero$, 
 for every $\xi_1>0$
  there exists a number $\eta>0$ such that 
$\bx \in N_{\eta} $ implies  
$\abs{D\bv_0(\bx)}< \xi_1$. 
For $\bb'_1, \bb'_2 \in N_{\eta}$,  since  $\bb'_1 + \theta \bbs \in N_{\eta}$, 
 (\ref{Y}) leads to 
\beq
  \abs{\bv_0 \(\bb'_1\) - \bv_0\(\bb'_2\)} < \xi_1 \abs{\bb'_1-\bb'_2}. 
\eeq

Next we consider estimation in a neighborhood of $\bc^*$. 
Putting $\bb= \ep \bb'$ and using the scaling property of $\bV$, 
we have 
\beqa
  \bV\(\bc^* + \bb\) &=& \ep^2  \bV\(\ba^* + \bb'\)
  \nn\\ 
  &=& \ep^2 \(  \bV\(\ba^*\) + D\bV\(\ba^*\)\bb' + \bv_0(\bb') \)
  \label{v}
\eeqa
Comparing (\ref{v0}) and (\ref{v}), we have 
\beqa
  \bv(\ep \bb' )  &=& \ep^2 \bv_0 \(\bb' \), 
\eeqa
which leads to  
\beqa
  \abs{ \bv(\bb_1 ) -  \bv(\bb_2 )} &=& \ep^2 \abs{ \bv_0 \(\bb'_1 \) - \bv_0 \(\bb'_2 \)}
  \nn\\ 
  &<&\ep^2 \xi_1 \abs{\bb_1' - \bb_2'}
  \nn\\
  &=& \ep \xi_1 \abs{\bb_1 - \bb_2}, 
\eeqa
for $\bb_1$, $\bb_2 \in N_{\ep \eta}$.  Repeating a similar argument for $\bef$, we get 
\beq
  \abs{ \bef(\bb_1 ) -  \bef(\bb_2 )}  <  \ep^2 \xi_2 \abs{\bb_1 - \bb_2}, 
\eeq
which implies (\ref{vandf2}) because $0<\ep <1$. 

\subsection{Lipschitz-type condition for ${\bH}(u, \bb)$}
Using (\ref{vandf1}) and (\ref{vandf2}), we readily obtain 
a Lipschitz-type condition for 
$\bH$ in the following form:   for any  $\xi >0$, 
  there exists $\eta >0$ such that  
\beq
  | \bH(u, \bb_1) - \bH(u, \bb_2) | <  \(\xi \ep + w \ep^{2} \) |\bb_1 - \bb_2|
  \label{lip}
\eeq
for all $\bb_1, \bb_2 \in N_{\ep \eta}$  and $0< \ep < 1$.  
Here, 
\beq
  w \equiv \abs{D\bF\(\ba^*\)}. 
\eeq

\noindent
{\it Proof}:
Recall 
\beq
  {\bH}(u, {\bb}) =  {\bv}({\bb}) + \rme^{-\ep u}\bF(\bc^{*}+{\bb}). 
\eeq
Then 
\beqa
  \abs{\bH\(u, \bb_1\) - \bH\(u, \bb_2\)} &\leq& \abs{\bv\(\bb_1\) - \bv\(\bb_2\)}
  + \rme^{-\ep u} \abs{D\bF\(\bc^*\) \(\bb_1 - \bb_2\)} 
  \nn\\
  &&+ 
   \rme^{-\ep u} \abs{\bef\(\bb_1\) - \bef\(\bb_2\)}
   \nn\\
  &\leq& \abs{\bv\(\bb_1\) - \bv\(\bb_2\)}
  + \ep^2 w \abs{\bb_1 - \bb_2} 
  \nn\\
 && + 
    \abs{\bef\(\bb_1\) - \bef\(\bb_2\)},   
\eeqa
where we have used  
\beq
  \abs{D\bF\(\bc^*\)} =  \ep^2  w. 
\eeq
Using  (\ref{vandf1}) and  (\ref{vandf2})  
for $\xi_1 = \xi_2 = \xi/2$, 
we get (\ref{lip}). 

\noindent
{\it Corollary:} if (\ref{lip}) holds, then 
 \beq
    \abs{\bH\(u, \bb_1\)}  <
  \( \xi \ep + \ep^2 w \) \abs{\bb_1}+ \rme^{-\al \ep u}\ep^3 \abs{\bF\(\ba^*\)}, 
   \label{estH}
 \eeq
 where $0< \al <1$.

 \noindent
{\it Proof}: 
Setting $\bb_2 \equiv \bzero$ in (\ref{lip}), we immediately get 
\beq
  \abs{\bH\(u, \bb_1\)- \rme^{-\ep u} \bF\(\bc^*\)}  < 
  \( \xi \ep + \ep^2 w \) \abs{\bb_1}, 
\eeq
which implies (\ref{estH}). 

\subsection{Estimation of $U$ and $T$}
Following  reference \cite{p} (Section 2.7), 
 we  derive:
there are positive constants $\al$, $\sigma$ 
and $K$,  such that 
\beqa
   | T(u) | < K \rme^{\sigma \ep u} &&(\mbox{for }u\leq 0) \nn\\
     |U(u)| < K \rme^{-(\al + \sigma) \ep u} && (\mbox{for } u \geq 0), 
     \label{TandU}
\eeqa
where 
$0<\al<1$.

\noindent
{\it Proof}: 
Let $ \mu_{j}\,  (j=1,..., n_+)$ be distinct eigenvalues 
of $Q$. Here 
$\Re \mu_j > 0$ for all $j$. We can choose sufficiently small $\sigma$ 
and sufficiently large $K$ 
such that 
\beqa
 &&\Re \mu_{j} >  \sigma > 0 \ \ \mbox{ for all $j=1,...,n_+$}
 \nn\\
&&  | T(u) | < K \rme^{\sigma \ep u} \ \ \mbox{for all $u\leq 0$.}
\label{T}
\eeqa

Similarly, let $\la_l\, (l=1,..., n_-)$ be distinct eigenvalues 
of $P$.
We take positive $\al'$ satisfying 
\beq
  \Re \la_l < - \al' < 0
  \label{al'}
\eeq
for all $l =1,...,n_-$.  Then there exists $K' > 0$ such that 
\beq
  |U(u)| < K' \rme^{- \al' \ep u}
\label{U}
\eeq
for all $u \geq 0$.  It should be noted that $\al'<1$ because of 
(\ref{-1}).  Further, if we choose $\si$ so small that 
\beq
  \Re \la_l < - \al' < -  \si < 0, 
\eeq
then we can write 
\beq
  \al' = \al + \si
\eeq
using some $0< \al <1$.
 Combining  (\ref{T}) and (\ref{U}),  we obtain (\ref{TandU}).

\subsection{Existence of a solution}
We define the space ${\cal C}$ of continuous mappings on $[0, \infty]$ into
$\bR^n$  in the following way:
\beq
  {\cal C} = \left\{ \bb; \abs{ \bb(u)} <  \ep \eta  \, \rme^{-\al \ep u},   \ u \geq 0\right\}. 
\eeq
   Define the metric on ${\cal C}$ by  
 \beq
  \rho\(\bb_1, \bb_2\) = \sup_{u \geq 0} \abs{\(\bb_1\(u\) -\bb_2\(u\) \)\rme^{\al\ep u}}, 
\eeq
for $\bb_1, \bb_2 \in {\cal C}$. 
Note that if $\bb \in {\cal C}$, $\bb(u) \in N_{\ep \eta}$ for all 
$u \geq 0$.  Then according to (\ref{lip}), for $\xi >0$, there exists $\eta >0$ such that 
\beqa
  \abs{ \bH\(u, \bb_1(u)\) -   \bH\(u, \bb_2(u)\) } &<&
   \(\xi \ep + w \ep^2 \) \abs{\bb_1(u) - \bb_2(u)} 
   \nn\\
   &<& \(\xi \ep + w \ep^2 \)\rho\(\bb_1, \bb_2\) \rme^{-\al \ep u}
   \label{rho2}
\eeqa
holds for $\bb_1, \bb_2 \in {\cal C}$, $u \geq 0$. 

Next we introduce the following mapping $\psi$ on ${\cal C}$:
 \beq
 \fl
   \psi\(\bb\)(u) = U(u) \bp + \int_{0}^{u}\rmd u' U(u-u') \bH(u', \bb(u')) 
   - \int_{u}^{\infty}\rmd u' T(u-u') \bH(u', \bb(u')). 
\label{psi}
 \eeq
 We have 
\beqa
  \abs{ \psi\(\bb_1\)(u)- \psi\(\bb_2\)(u)} &\leq& 
  K\(\xi \ep  + \ep w^2\) \rho\(\bb_1, \bb_2\) 
  \nn\\
  &&\times
  \( \int_{0}^u \rme^{-(\al+\sigma)\ep (u-u')} \rme^{-\al \ep u'}  \rmd u'  \right. 
  \nn\\
  &&+ 
  \left. \int_u^\infty  \rme^{\sigma(u-u')\ep }  \rme^{-\al \ep u'} \rmd u'  \), 
\eeqa 
according to (\ref{TandU}) and (\ref{rho2}).  
Here the first integration is evaluated as 
\beqa
  \int_{0}^u \rme^{-(\al+\sigma)\ep (u-u')} \rme^{-\al \ep u'}  \rmd u' 
  = \frac{1}{\si \ep }\( \rme^{-\al \ep u}- \rme^{-\(\al + \si\)\ep u} \)
 <  \frac{\rme^{-\al \ep u}}{\si \ep }. 
 \label{int1}
\eeqa
The second one is 
\beq
   \int_u^\infty  \rme^{\sigma(u-u')\ep }  \rme^{-\al \ep u'} \rmd u' 
   = \frac{1}{(\al +\sigma)\ep} \rme^{-\al \ep u} 
   < \frac{\rme^{-\al \ep   u}}{\si \ep}. 
   \label{int2}
\eeq
In this way,  we obtain 
\beq
  \rho\(\psi\(\bb_1\), \psi\(\bb_2\) \) \leq 
  \frac{2 K}{\si}\(\xi   + \ep w \) \rho\(\bb_1, \bb_2\), 
\label{rhopsi}
\eeq
for  $\bb_1, \bb_2 \in {\cal C}$. 

Now we show that $\psi$ is a contraction map on  ${\cal C}$ 
into itself 
if  we  choose $\xi$, $\eta$, $\ep$ and $\bp$ such that 
the following inequalities hold:
\beqa
  K  \( \abs{\bp} 
  +\frac{2}{\si}  \ep^2 \abs{\bF\(\ba^*\)}\) &<& \frac{1}{2} \ep \eta
  \label{cond1}
  \\
   \frac{2 K}{\si} \(\xi  + \ep w \) &<& \frac{1}{2}. 
  \label{cond2}
\eeqa
It is easy to see that such choice is in fact possible.\footnote{
For example, 
 we first fix $\xi = \xi_0$ such that  $0< \xi_0 < \si/(4K)$, which is realized 
 by an appropriate choice of $\eta(=\eta_0)$ according to (\ref{lip}). 
  Next we choose 
$\ep=\ep_0$ satisfying 
$$
  \ep_0 < \min\(\frac{\si \eta_0}{4 K \abs{\bF\(\ba^*\)}}, \frac{1}{w} \( \frac{\si}{4K} - \xi_0 \)\). 
$$
By construction,  $(\ep_0, \xi_0)$ satisfies (\ref{cond2}). 
Furthermore,  choosing $\bp$ such that 
$$
0<  \abs{\bp}  < \frac{2 \ep_0 \abs{\bF\(\ba^*\)}}{\si}
\( \frac{\si \eta_0}{4 K \abs{\bF\(\ba^*\)}} - \ep_0\), 
$$
we find that (\ref{cond1}) holds. 
}

When (\ref{cond1}) and (\ref{cond2}) hold, we can show that 
$\psi$ maps ${\cal C}$ into itself.  Let us recall (\ref{estH}). 
If ${\bb} \in {\cal C}$, then 
\beq
  \abs{{\bH}\(u, {\bb}(u)\)}   < 
  \( \ep \eta \( \xi \ep + \ep^2 w \) +  \ep^3
  \abs{{\bF}\(\ba^*\)} \) \rme^{-\al \ep u}. 
\label{estH2}
\eeq
 Combining this and the estimations (\ref{int1}) and (\ref{int2}), 
 (\ref{psi}) yields
\beqa
  \abs{\psi\(\bb \)(u)} &\leq& K \rme^{-(\al + \si) \ep u} \abs{\bp} 
   +   \( \ep \eta \( \xi  + \ep w \) +  \ep^2
  \abs{{\bF}\(\ba^*\)} \) \frac{2K}{\si} \rme^{-\al \ep u}
  \nn\\
  &<&  K \( \abs{\bp} +\frac{2K}{\si} \ep^2 \abs{{\bF}\(\ba^*\)}\)\rme^{-\al \ep u}
   +  \ep \eta \( \xi  + \ep w \)  \frac{2K}{\si} \rme^{-\al \ep u}
   \nn\\
   &<& \ep \eta  \rme^{-\al \ep u}
\eeqa
for all $u \geq 0$.  In the last inequality, we have used (\ref{cond1}) and 
(\ref{cond2}).  
 It shows that $\psi\(\bb\) \in {\cal C}$. 
 Furthermore, 
 (\ref{rhopsi}) and (\ref{cond2}) indicate  that 
$\psi$ is a contraction map.

Thus, existence ${\bb} \in {\cal C}$ of the solution
\beq
  {\bb} = \psi\({\bb}\) 
\eeq
and its uniqueness follow from the fixed point theorem
(see the textbook  \cite{kf}, for example). 
Since ${\bb}$ belongs to ${\cal C}$, we find that there is a positive number 
$J$ such that 
\beq
  \abs{{\bb}(u)} < J  \rme^{-\al \ep u}
\eeq
for $u \geq 0$.

  \section{Estimation of the remaining terms in (\ref{ie})}
  \label{appc}
According to (\ref{ie}) and (\ref{ub})-(\ref{intl}), 
  it is sufficient to show that 
\beqa
 &&\abs{\int_0^u U(u-u') \( \bH(u', \tilde{\bb}(u')) - \rme^{-\ep u}
  \tilde{\bF}\(\tilde{\bc}^*\)\)\rmd u' } < \sum_{l=0}^{n_-}G^{(l)}_3(u) \rme^{\la_l \ep u}
\label{estrem1}
\\
 &&\abs{\int_{u}^{\infty} T(u-u') \bH(u', \tilde{\bb}(u')) \rmd u'} < B \rme^{- \ep u}, 
  \label{estrem2}
\eeqa
for all $u \geq 0$, in order to complete the derivation of 
 (\ref{premain}).  Here $B$ is a positive constant and 
 $G_3^{(l)}(u)$ is a polynomial whose degree is 
at most $\nu_l -1$.  

Let us first show (\ref{estrem1}). 
According to (\ref{f}),  $\tilde{\bH}$ is written as 
\beq
  \tilde{\bH}\(u, \tilde{\bb}(u)\) = \tilde{\bv}\(\tilde{\bb}(u)\) + 
  \rme^{-\ep u} \( \tilde{\bF}\(\tilde{\bc}^*\) +
  D\tilde{\bF}\(\tilde{\bc}^*\)\tilde{\bb}(u) 
  + \tilde{\bef}\(\tilde{\bb}(u)\) \). 
\eeq
Since $\tilde{\bV}, \tilde{\bF} \in C^2(E)$, \footnote{
see the second footnote in Section 2.
}  the reminders $\tilde{\bv}, \tilde{\bef} $ also 
belong to $C^2(E)$.  Using the Taylor theorem, we find that there exists a number 
$\theta$ ($0 \leq \theta \leq 1$) such that 
\beq
  \tilde{v}_i \(\tilde{\bb}\) = \sum_{j k}\frac{1}{2} 
  \frac{\del^2 \tilde{v}_i}{\del \tilde{b}_j \del \tilde{b}_k} 
  ( \theta \tilde{\bb}) \tilde{b}_j \tilde{b}_k
\eeq
holds.  
Employing the estimation (\ref{estb}) for $\tilde{\bb}(u)$,  
we can show that for sufficiently large $B_1$
\beq
    \abs{ \tilde{\bv} \(\tilde{\bb}(u)\)} <  B_1 \rme^{-2\al \ep u}. 
\label{bv}
\eeq
A similar estimation holds true for $\tilde{\bef}$, i.e.,  for some $B_2 > 0$, 
\beq
    \abs{ \tilde{\bef} \(\tilde{\bb}(u)\)} <  B_2 \rme^{-2\al \ep u}.
    \label{bf}
\eeq

If the assumption (\ref{ess}) holds, 
 we can take 
 \beq
   \frac{1}{2} < \al < 1. 
   \label{1/2}
 \eeq
 Further, we  choose $\al$ such that 
 $2\al \neq -\la_l$ for all $l = 1, ..., n_-$ for later use. 
Combining the condition (\ref{1/2}) with (\ref{bv}) and (\ref{bf}), 
it follows that there is a number 
 $B_3 >0$ such that  
\beqa
\fl
   \abs{\tilde{\bH}\(u, \tilde{\bb}(u)\)-
   \rme^{-\ep u} \tilde{\bF}\(\tilde{\bc}^*\)} &<&  
   \abs{\tilde{\bv}\(\tilde{\bb}(u)\)} + 
  \rme^{-\ep u}\( 
   \abs{D\tilde{\bF}\(\tilde{\bc}^*\)\tilde{\bb}(u) }
  + \abs{\tilde{\bef}\(\tilde{\bb}(u)\)} \)
  \nn\\
  &<& B_1 \rme^{-2 \al \ep u} +  \abs{D\tilde{\bF}\(\tilde{\bc}^*\)}\rme^{-(1+ \al)\ep u} +
  B_2 \rme^{(-1-2 \al) \ep u}
  \nn\\
  &<& B_3 \rme^{-2\al \ep u}
  \label{b8}
\eeqa
for all $u \geq 0$. 
Since the same estimation holds for every component projected onto $W_l$, 
we obtain 
\beqa
 &&\abs{U(u-u') \( \tilde{\bH}(u', \tilde{\bb}(u')) - \rme^{-\ep u}
  \tilde{\bF}\(\tilde{\bc}^*\)\) } 
  \nn\\
  &=& 
  \abs{\sum_{l=1}^{n_-} \rme^{\la_l \ep (u-u')} \sum_{k=0}^{\nu_l -1} \frac{\(\ep (u-u')\)^kN_l^k}{k!}
  \(\tilde{\bH}^{(l)}(u', \tilde{\bb}(u')) - \rme^{-\ep u}
  \tilde{\bF}^{(l)} \(\tilde{\bc}^*\)\)}
  \nn\\
  &<&
  B_4  \sum_{l=1}^{n_-} \rme^{\la_l \ep (u-u')} \sum_{k=0}^{\nu_l -1} \frac{\(\ep (u-u')\)^k}{k!}
  \rme^{-2 \al \ep u'}. 
  \label{c8}
\eeqa
Integrating the both sides by $u'$ over $[0,u]$,  we obtain (\ref{estrem1}). 

As for  (\ref{estrem2}), we use (\ref{b8}). Then we have 
\beq
  \abs{\tilde{\bH}\(u', \tilde{\bb}(u')\)} < 
 \abs{\tilde{\bF}(\tilde{\bc^*})}  \rme^{-\ep u'} + B_3 \rme^{-2\al \ep u'}
  < C \rme^{-\ep u'}
\eeq
for some $C>0$ because we have set $\al > 1/2$.  
Then the integral is easily estimated as 
\beq
\fl
  \abs{\int_u^\infty \rmd u' T(u-u') \tilde{\bH}\(u', \tilde{\bb}(u') \) }< 
  K C  \int_u^\infty \rmd u' \rme^{\si \ep \(u-u'\)} \rme^{-\ep u'} 
  = \frac{KC}{(\si +1) \ep} \rme^{-\ep u}. 
\eeq
Thus we get (\ref{estrem2}).

\end{document}